\documentclass[final,3p,times,twocolumn]{elsarticle}

\usepackage{amsmath}
\usepackage{graphicx}
\usepackage{dcolumn}
\usepackage{bm}
\usepackage{multirow}
\usepackage{float}
\usepackage{placeins}
\usepackage{siunitx}
\usepackage{amssymb}
\usepackage{booktabs}
\usepackage{xcolor}
\usepackage[colorlinks=true,linkcolor=blue,citecolor=blue,urlcolor=blue]{hyperref}
\usepackage{miller}
\usepackage{silence}
\usepackage{tabularx}
\usepackage{orcidlink}

\journal{}

\title{SPARSE - Efficient High-Resolution SEM Imaging of Rare Microstructural Features Across Large Areas by Selective Rescanning}

\author[IMM]{Tom Reclik\orcidlink{0009-0007-9395-5912}}
\ead{reclik@imm.rwth-aachen.de}

\affiliation[IMM]{organization={Institute of Physical Metallurgy and Materials Physics, RWTH Aachen University},
            city={Aachen},
            postcode={52074},
            country={Germany}}

\author[TUD]{Jan Gerlach\orcidlink{0000-0003-0429-0204}}

\author[IMM]{Maximilian A. Wollenweber\orcidlink{0009-0003-0921-5382}}

\affiliation[TUD]{organization={Institute of Forming Technology and Lightweight Components, TU Dortmund}, city={Dortmund}, postcode={44227}, country={Germany} }

\author[TUD]{Yannis P. Korkolis\orcidlink{0000-0001-9482-9394}}

\author[IMM]{Sandra Korte-Kerzel\orcidlink{0000-0002-4143-5129}}

\author[IMM]{Ulrich Kerzel\orcidlink{0000-0002-4939-6726}}

\begin{document}

\begin{abstract}
Characterisation of rare microstructural features in scanning electron microscopy (SEM) requires imaging large areas at high resolution. This leads to prohibitively long acquisition times. We present an open-source Python framework that addresses this bottleneck through a two-stage approach: a fast scan identifies regions of interest, which are then selectively rescanned with imaging parameters suitable for quantitative analysis. The framework defines a generic microscope interface and a modular detection interface, allowing adaptation to different microscope platforms and detection methods. Scanning, detection, and rescanning are parallelized using separate processes, ensuring that computation time does not extend acquisition time. The two processes communicate exclusively through queues, avoiding shared mutable state and eliminating the need for explicit synchronization. We validate the framework on damage detection in dual-phase DP800 steel using a Tescan Clara SEM. For a representative configuration a detection rate of \SI{99}{\percent} is achieved at approximately \SI{58}{\percent} of the conventional acquisition time. At \SI{95}{\percent} detection rate, acquisition time drops to \SI{19}{\percent}. These time savings estimates represent lower bounds based on the ratio of scanned pixels. The complete implementation will be made available upon publication and upon request during peer-review.

\end{abstract}

\maketitle

\section{Introduction}

Characterising rare microstructural features such as damage sites in dual-phase steels \cite{tasan_overview_2015, kusche_large-area_2019} or non-metallic inclusions in clean steels \cite{atkinson_characterization_2003} requires surveying large sample areas at high spatial resolution. Conventional approaches to this problem, whether manual inspection or automated panoramic imaging \cite{kusche_large-area_2019}, treat every part of the sample equally, acquiring high-resolution data across the entire region of interest. Yet for rare features, such as damage in dual-phase steels or inclusions in 16MnCrS5 case-hardening steel \cite{wollenweber_automated_2023}, the vast majority of the imaged area contains no relevant information. This creates a fundamental inefficiency: acquisition time scales with the total surveyed area rather than with the amount of relevant information, making comprehensive studies prohibitively slow.

In prior work, Dahmen et al.\ showed that adaptive pixel-dwell sampling based on initial low-dose images can substantially reduce acquisition time \cite{dahmen_feature_2016}, and Clusiau et al.\ demonstrated feature tracking to image only regions of interest at high resolution \cite{clusiau_workflow_2025}. These approaches demonstrate the potential of selective imaging strategies, but differ in how candidates are identified and how the imaging resources are allocated. When the initial scan is performed at reduced quality to accelerate acquisition, a trade-off arises between detection completeness and efficiency: more aggressive parameter reduction increases time savings but risks missing small or low-contrast features. Existing approaches have not systematically characterised this trade-off, and most implementations are tightly coupled to specific hardware or detection methods, limiting their transferability across platforms and applications.

\begin{figure*}[!ht]
    \centering
    \includegraphics[width=\linewidth]{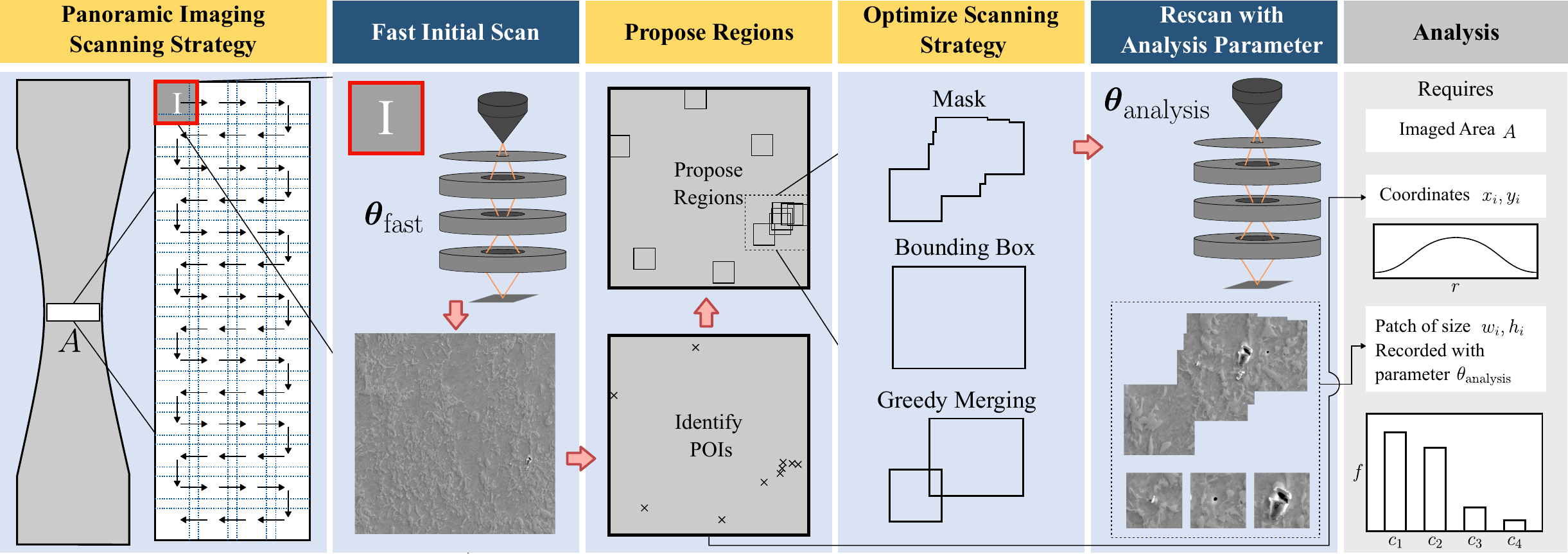}
    \caption{Overview of the automated scanning framework SPARSE. From left to right: the panoramic imaging strategy defines the tile grid covering the target area $A$. Each tile is first acquired with fast imaging parameters $\boldsymbol{\theta}_{\text{fast}}$. From the fast scan, POIs are identified and regions are proposed around detected candidates. The proposed regions are then optimized into a scanning strategy. Two approaches are available depending on microscope capabilities: bounding box extraction for systems limited to rectangular scans, or greedy merging of a binary mask for systems supporting arbitrary scan regions. Selected regions are rescanned with analysis parameters $\boldsymbol{\theta}_{\text{analysis}}$. Finally, the analysis receives the coordinates of each rescan patch relative to the panorama and the corresponding high-resolution images. Colour coding indicates operations performed by the microscope (blue) versus the framework software (yellow), with downstream analysis shown in grey.}
    \label{fig:workflow}
\end{figure*}

In this work, we present an open-source framework for efficient large-area SEM characterization of rare features (SPARSE - Selective Parallelized Adaptive Rescanning for SEM Efficiency) illustrated in \autoref{fig:workflow}. 
The framework is built on two principles: modularity and parallelization. Detection methods and microscope control are implemented as interchangeable components; users implement a microscope interface class providing the required methods for their specific hardware, and a detection function appropriate for their application. This separates the open-source scanning logic from vendor-specific microscope control. Scanning, detection, and rescanning are parallelized to maximize microscope utilization, eliminating idle time during image analysis. The framework is implemented as a Python library, allowing integration into existing Python-based analysis workflows.  Users define their scanning task through a configuration specifying the survey area, resolution tiers, and detection parameters. Starting from provided examples, the default scripts cover common tasks such as surveying large rectangular regions. The primary user decision lies in configuring the detection pipeline: how aggressively to reduce acquisition parameters, such as dwell time or resolution, for the initial fast scan, and which detection method to apply to the resulting images. For parametric detection methods, a sweep mode allows exploration of the accuracy–efficiency trade-off before committing to a full acquisition. An example implementation for TESCAN instruments via SharkSEM is provided.

The framework's performance is validated through a case study on microstructural damage characterization in dual-phase steel. A well-investigated research area, that leverages statistical information on rare, heterogeneously distributed microstructural voids to increase component performance \cite{tekkaya_forming-induced_2017, hering_damage-induced_2020, medghalchi_damage_2020,hannard_characterization_2016}. In the present case, we apply the framework to a dual-phase steel, where damage sites typically require both high-resolution and large area imaging for accurate quantification\cite{avramovic-cingara_effect_2009, asik_microscopic_2019, wollenweber_automated_2023}. We demonstrate substantial time savings compared to conventional previously applied (\cite{kusche_large-area_2019,medghalchi_damage_2020}) full-area high-resolution imaging and characterize the trade-off between detection completeness and acquisition efficiency. The complete implementation is available as open-source software.

\section{Framework Architecture}

The following subsections describe the terminology (section~\ref{sec:Nomenclature}), design principles (section~\ref{sec:DesignPrinciples}), user workflow (section~\ref{sec:UserWorkflow}), scanning procedure (section~\ref{sec:ScanningProcedure}), and implementation details (section~\ref{sec:Implementation}).

\subsection{Nomenclature}
\label{sec:Nomenclature}
Throughout this work, we use the following terminology:
\begin{itemize}
    \item \textit{Panorama}: The complete tiled area on the sample to be scanned, covering an area too large for manual image-by-image acquisition.
    \item \textit{Tile}: A single field of view within the panorama.
    \item \textit{Partition}: A subdivision of each tile into equally sized sections to enable parallel workflows. The number of partitions is configurable; in this work, each tile is divided into four parts.
    \item \textit{POI}: Point of interest; a candidate location that may contain a feature. Whether a POI corresponds to an actual feature of interest or to an artifact is determined by the subsequent analysis and is independent of the framework.
    \item \textit{Region}: Region of interest; a region around one or more POIs to be rescanned at higher quality.
    \item \textit{$\boldsymbol{\theta}_{\text{fast}}$}: Vector of imaging parameters for the fast initial scan, such as resolution, dwell time, working distance, acceleration voltage, and beam current. These are chosen to accelerate acquisition while retaining sufficient image quality for POI identification.
    \item \textit{$\boldsymbol{\theta}_{\text{analysis}}$}: Vector of imaging parameters required for subsequent quantitative analysis. These define the target image quality and are determined by the requirements of the downstream analysis.
\end{itemize}

\subsection{Design Principles}
\label{sec:DesignPrinciples}
The framework is built on two core principles: Modularity and Parallelization.

\textit{Modularity.} The framework separates microscope control from the scanning logic through a generic microscope interface. Users provide an implementation of this interface for their specific hardware, allowing the framework to be adapted to different microscope models. Similarly, detection algorithms are interchangeable components: users implement a detection function through the provided interface, choosing an approach appropriate for their application. An example implementation using thresholding and DBSCAN clustering is provided with the framework. Additional functionality for large-area scanning—such as failure recovery, multi-session scanning, and parameter drift compensation—is available as optional modules, which can be further expanded. The specific architectural choices and their rationale are described in section~\ref{sec:Implementation}.

\textit{Parallelization.} Microscope time is the most valuable resource. The framework offloads POI identification and region calculation to a separate process, keeping the microscope continuously occupied. The efficiency of this approach depends on the native feature density and the speed of POI identification. 

\subsection{User Workflow}
\label{sec:UserWorkflow}

To apply the framework to a new use case, users require two components: a microscope interface for their hardware providing the required methods described above, and a detection function that receives a fast scan image and returns regions that are to be rescanned. Before scanning, it is recommended to evaluate the detection method on recorded image pairs acquired with $\boldsymbol{\theta}_{\text{fast}}$ and $\boldsymbol{\theta}_{\text{analysis}}$ to verify that POIs are reliably identified in the fast scan images and to determine suitable fast scan parameters. The framework provides a mode to automatically acquire these paired images, configured through the same configuration file as used for the full scan.

With these components in place, the user configures the scan and panorama parameters in a YAML configuration file and starts the scan. The interactive setup phase is shown in \autoref{fig:flow_chart}: the user focuses and adjusts brightness and contrast at three corner positions of the panorama, from which the framework interpolates imaging parameters across the scan area. Three corners are sufficient to define a linear gradient across the panorama, which assumes an approximately planar sample surface with possible tilt. Once the setup is complete, the framework proceeds autonomously through the parallelized scanning workflow.
\begin{figure}[!ht]
    \centering
    \includegraphics[width=0.8\linewidth]{figures/flow_chart.pdf}
    \caption{Flowchart of the scanning procedure. Blue boxes indicate steps requiring user interaction; green boxes indicate automated processes. Users set imaging parameters at three corner positions, from which the framework interpolates parameters across the entire panorama.}
    \label{fig:flow_chart}
\end{figure}

\subsection{Scanning Procedure}
\label{sec:ScanningProcedure}
Efficient imaging requires the microscope to scan arbitrary sub-regions of the field of view, as stage movements between scans add overhead and introduce misalignment. Each tile is divided into four partitions, and for each partition the process consists of: (1) fast initial scan, (2) POI identification, (3) region proposal, and (4) analysis rescan. 

The number of partitions balances two competing effects: more partitions provide more time for the detection process, as regions from earlier partitions can be rescanned while later partitions are still being acquired. However, features that lie on a partition boundary may be split across adjacent partitions and missed by the detection. The probability of this occurring scales with the number of boundary lines and quadratically with the minimum feature size, but inversely quadratically with the tile size in pixels. For the tile and feature sizes used in this work, this probability is negligible. Two partitions would minimize boundary effects but provide limited opportunity for overlapping detection and rescanning with acquisition: the detection must complete and all rescan regions must be acquired before the second partition finishes scanning. With four partitions, the SEM can rescan detected regions from earlier partitions while later partitions are still being acquired and processed. This balances detection time against boundary effects.

\begin{figure*}[!ht]
    \centering
    \includegraphics[width=0.8\linewidth]{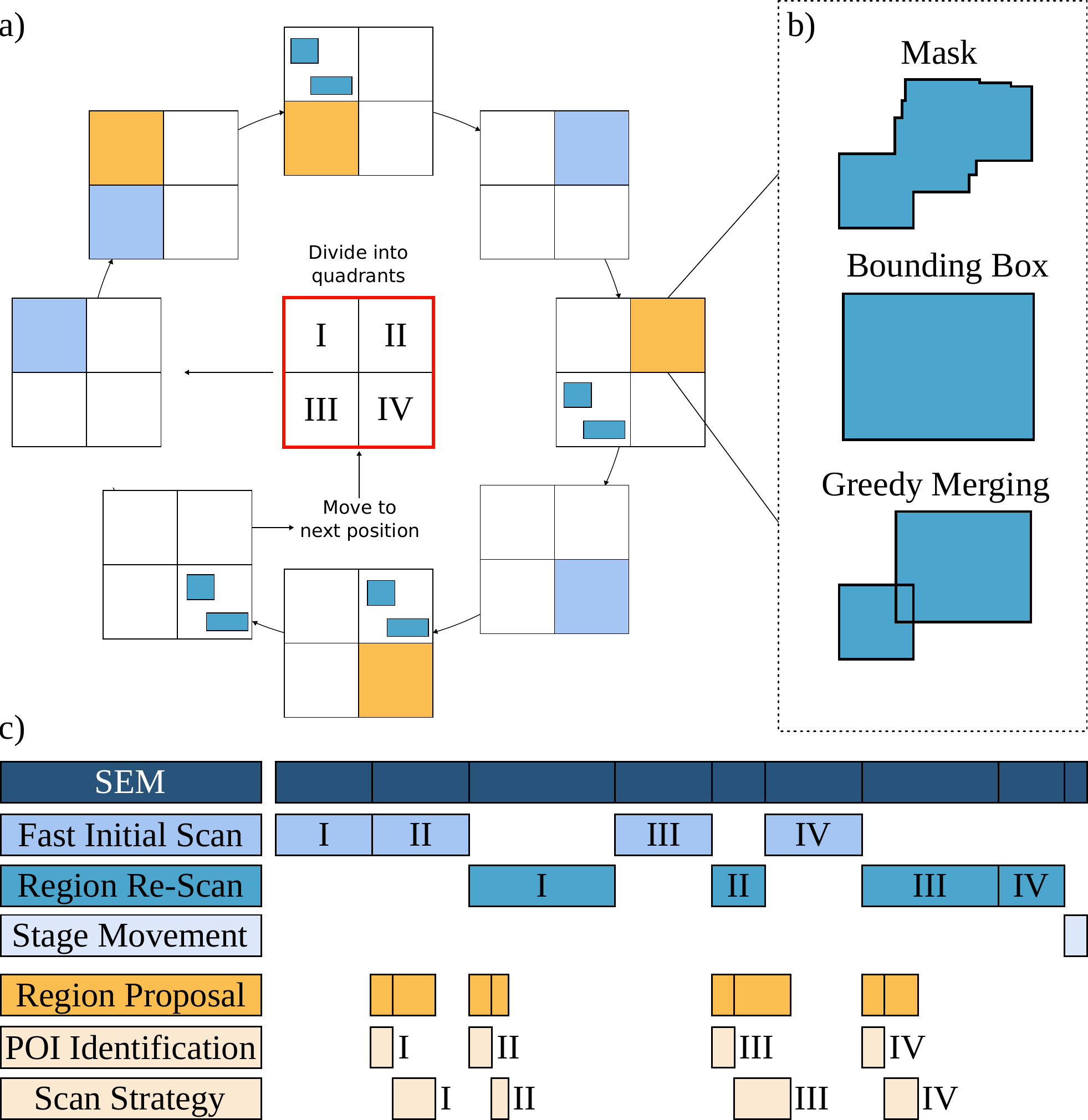}
    \caption{Parallelized workflow for a single tile divided into four partitions. Blue indicates SEM operations (fast scanning or rescanning); orange indicates concurrent analysis on the workstation. While the microscope performs the fast scan of partition $n$, region proposal for partition $n-1$ proceeds in parallel. Once the fast scan completes, the microscope rescans detected regions in partition $n-1$ before advancing to the fast scan of partition $n+1$. This pattern ensures analysis time is hidden behind acquisition time.}
    \label{fig:parallel_workflow}
\end{figure*}

These four steps of the scanning procedure are parallelized as shown in \autoref{fig:parallel_workflow}. Initially, the microscope scans the first two partitions with $\boldsymbol{\theta}_{\text{fast}}$. Once the first partition is acquired, the detection process begins analysing it while the microscope continues with the second partition. After the second scan completes, region proposals for the first partition should be ready, allowing the microscope to re-scan the first partition while analysis of the second partition proceeds. This alternating pattern continues until all regions in the final partition have been re-scanned. Then the microscope moves to the next tile. The following subsections describe each step in detail.

\subsubsection{Fast Initial Scan}
The first scan is performed using imaging parameters $\boldsymbol{\theta}_{\text{fast}}$ that result in a faster initial acquisition. The resulting image is used to localise points of interest. The specific imaging parameter choice needs to balance an acceleration in imaging with enough recorded information to still be able to localise points of interest. The fast scan parameters can employ a lower resolution, reduced dwell time, or both.

\subsubsection{POI Identification}
From the fast scan image, points of interest are identified using the selected detection method. This function receives the fast scan image as a NumPy array and returns a list of POI coordinates (see section~\ref{sec:DetectionInterface} for details).  The framework is agnostic to the specific algorithm: threshold-based segmentation \cite{zhou_efficientand_2024} or deep learning approaches \cite{wollenweber_automated_2023, shah_automated_2023} can all be used through the same interface.

\subsubsection{Region Proposal}
\label{sec:RegionProposal}
Given the identified POIs, rectangular regions of size $w_{\text{region}} \times h_{\text{region}}$ are generated around each point. Since these regions may overlap, scanning each separately would result in redundant acquisition. The framework provides three strategies for combining overlapping regions, and the user selects the appropriate one in the configuration based on their microscope's scanning capabilities. The choice depends on the scanning time per region $t_{\text{img}}$, which is given by:
\begin{equation}
    t_{\text{img}} = t_{\text{dwell}} \cdot n_{\text{pixel}} + t_{\text{row}} \cdot n_{\text{rows}} + t_{\text{overhead}}
\end{equation}
where $t_{\text{dwell}}$ is the dwell time per pixel, $n_{\text{pixel}}$ is the total number of pixels, $t_{\text{row}}$ is the line flyback time, $n_{\text{rows}}$ is the number of scan lines, and $t_{\text{overhead}}$ is the fixed overhead between scans.

\textit{Mask-based merging.} If the microscope's scan generator accepts arbitrary masks, all regions are combined into a single binary mask. The user provides the conversion from this mask to the format required by their scan generator as part of the microscope interface. This is the most efficient strategy when available.

\textit{Bounding box merging.} Overlapping regions are merged into their combined bounding box. This is efficient when $t_{\text{overhead}}$ is large relative to the cost of scanning additional pixels. This is the simplest strategy.

\textit{Greedy merging.} For each pair of overlapping regions, the times for separate scanning $t_{\text{separate}}$ versus merged scanning $t_{\text{merge}} $ compare as follows:
\begin{equation}
    t_{\text{separate}} = t_{\text{dwell}}(n_1 + n_2) + t_{\text{row}}(n_{1,\text{rows}} + n_{2,\text{rows}}) + 2t_{\text{overhead}}
\end{equation}
\begin{equation}
    t_{\text{merge}} = t_{\text{dwell}} \cdot n_{\text{merged}} + t_{\text{row}} \cdot n_{\text{merged,rows}} + t_{\text{overhead}}
\end{equation}
where $n_{\text{merged}}$ and $n_{\text{merged,rows}}$ refer to the merged bounding box dimensions. Regions are merged only if $t_{\text{merge}} < t_{\text{separate}}$. This strategy is most efficient when $t_{\text{overhead}}$ is small.

\subsubsection{Region Rescan}
\label{sec:rescanning}
The optimized regions are scanned using $\boldsymbol{\theta}_{\text{analysis}}$, the imaging parameters required for subsequent quantitative analysis.

\subsection{Implementation Details}
\label{sec:Implementation}

The framework is implemented in Python 3.11 and runs on a workstation connected to the microscope control PC over the network. The reference implementation communicates with the microscope through the Tescan SharkSEM API. The framework does not depend on any platform-specific libraries apart from the vendor-specific microscope control interface.

The framework is implemented as a Python library using abstract base classes to define the microscope and detection interfaces. Users subclass these and pass instances to the framework at initialization, for example \texttt{sem = TescanClara(params)}, where \texttt{TescanClara} inherits from the framework's base microscope class. This approach was chosen over alternative architectures for its simplicity. A microservices architecture, where microscope control, detection, and scan planning run as separate services communicating over a network, would allow each component to run on a different machine and in a different language. However, it introduces substantial complexity in deployment, error handling, and inter-service communication that is not justified for a two-process pipeline where low latency between microscope commands is essential. Asynchronous frameworks such as asyncio\footnote{\url{https://docs.python.org/3/library/asyncio.html}} operate within a single thread and cannot provide true parallelism for CPU-bound detection tasks. Distributed computing frameworks such as Dask\cite{rocklin_dask_2015} would allow scaling detection across multiple machines but add a heavy dependency and scheduling overhead that does not pay off for the current workload. The library approach keeps the barrier to adoption low for researchers familiar with the scientific Python ecosystem, while the abstract base class pattern ensures that new hardware integrations follow a consistent structure.

\subsubsection{Parallelization Strategy}
\label{sec:ParallelizationStrategy}
The parallelized workflow described in section \ref{sec:ScanningProcedure} is realized using Python's \texttt{multiprocessing} standard library with two processes, the SEM process and the region proposal process, communicating through two shared queues. The SEM process holds the full partition sequence and advances through it autonomously. After acquiring a partition with $\boldsymbol{\theta}_{\text{fast}}$, it passes the image to the region proposal queue and checks its own queue for pending rescan regions. If rescan regions are available, they are scanned before advancing to the next partition. If not, the SEM process proceeds to the next partition immediately. The region proposal process receives images, performs POI identification and region proposal, and adds detected regions to the SEM queue for rescanning. Since the SEM process manages its own scanning sequence, it is never idle waiting for the region proposal process to complete. Both partition scans and rescan regions use the same coordinate format, so the SEM process handles them identically. Inter-process communication via queues requires serialization of the transmitted data. The SEM process passes acquired partition images as NumPy arrays to the region proposal queue, while the region proposal process sends scan coordinates back to the SEM queue. Rescanned regions are saved directly to disk by the SEM process and are not passed between processes. The serialization overhead scales with image size but remains negligible relative to acquisition time: even for a $3072 \times 3072$ pixel 16-bit image (\SI{18}{\mega\byte}), serialization takes on the order of milliseconds, while acquiring such an image at typical dwell times takes seconds to minutes. For applications where this overhead becomes relevant, Python's \texttt{multiprocessing.shared\_memory}\footnote{\url{https://docs.python.org/3/library/multiprocessing.shared_memory.html}} module could be used to avoid copying.

Process-based parallelism is chosen over thread-based concurrency to ensure that the SEM process is never blocked by the region proposal process. In Python, threads share a global interpreter lock (GIL), which prevents concurrent execution of Python code. While many numerical libraries release the GIL during computation, the framework cannot assume this for user-provided detection methods. Separate processes each operate with an independent interpreter, ensuring that a computationally intensive detection step cannot block microscope communication. Recent Python versions (3.13+) introduce experimental free-threaded builds that remove the GIL, which would allow thread-based parallelism for CPU-bound tasks. However, as this feature is not yet the default and library support remains incomplete, the framework uses process-based parallelism to ensure compatibility across standard Python installations. Since the two processes do not share mutable state and communicate exclusively through thread-safe queues, no explicit synchronization mechanisms such as mutex locks are required, and race conditions are avoided by design.

The main process monitors the region proposal process. If it terminates unexpectedly, the main process triggers the shutdown procedure. Queue operations use timeouts to prevent indefinite blocking in case communication is interrupted. During shutdown, whether due to an error or successful completion, the controller writes the index of the last completed tile to disk. Upon restart, the framework reads this index and resumes scanning from the next tile, avoiding the need to repeat already acquired data.

\subsubsection{Coordinate systems}
\label{sec:CoordinateSystem}
Stage movements between tiles use the microscope coordinate system in millimetres. Within a tile, all operations, including detection, region proposal, and rescanning, use pixel coordinates relative to the field of view. Rescanning detected regions does not require stage movement, as the microscope scans arbitrary sub-regions of the field of view by directing the electron beam (see \texttt{scan\_area} in the microscope interface).

\subsubsection{Microscope Interface}
The framework defines an abstract base class for microscope control. To adapt the framework to a new microscope, users inherit from this class and map the required methods to the corresponding functions in their microscope's control software or API. The implemented class is then passed to the framework at initialization, which handles integration with the scanning controller.
The following methods need to be implemented:

\begin{itemize}
    \item \texttt{connect()}: Establish a connection to the microscope. The method is expected to raise an exception if the connection fails.
    \item \texttt{disconnect()}: Close the connection to the microscope.
    \item \texttt{prepare()}: Perform preparation steps such as enabling the electron beam and selecting the detector. The specific steps depend on the instrument and are determined by the user. The framework calls this method once before the scanning procedure begins and stores the initial stage position, imaging parameters, and working distance for later restoration.
    \item \texttt{scan\_area(coordinates, dwell\_time, window\_size)}: Scan a specified sub-region of the field of view and return the image as NumPy array. The window size and scan window coordinates are specified in pixels, the dwell time in nanoseconds. This differs from scanning the full FOV and is essential for both the parallelization strategy and targeted rescanning of detected regions (see section~\ref{sec:rescanning}).
    \item \texttt{move\_stage(coordinates)}: Move the stage to the specified position. Coordinates are in millimetres in the microscope coordinate system (see section~\ref{sec:CoordinateSystem}.
    \item \texttt{reset\_microscope()}: Return the stage to its initial position and restore all imaging parameters. If configured, the electron beam is switched off. This method is also called in case of an error to ensure the microscope is left in a defined state.
    \item \texttt{save\_metadata(path)}: Save the current stage position, device parameters, and timestamp to the specified path.
\end{itemize}

Optional methods include:
\begin{itemize}
    \item \texttt{set\_working\_distance(value)}: Adjust focus. The working distance is specified in millimetres. Required for large-area scans where sample tilt causes focal drift.
    \item \texttt{get\_working\_distance()}: Retrieve the current working distance in millimetres. Required to determine working distance gradient for large-area scans.
    \item \texttt{set\_brightness\_contrast(brightness, contrast)}: Adjust detector settings. Values are specified in percent. Required when imaging conditions vary across the scan area.
    \item \texttt{get\_brightness\_contrast()}: Get brightness and contrast values in percent to estimate brightness and contrast gradients over large-area scans.
\end{itemize}

\subsubsection{Detection Interface}
\label{sec:DetectionInterface}

The detection method is implemented as a function that receives the fast scan image as a NumPy array and returns a list of POI coordinates in pixels relative to the field of view at the analysis resolution defined by $\boldsymbol{\theta}_{\text{analysis}}$. The framework then generates regions around the identified POIs and combines overlapping regions using the merging strategies described in section~\ref{sec:RegionProposal}. The region dimensions are specified in the configuration file and depend on the requirements of the subsequent analysis, for example the input size of a classification model.

\subsubsection{Additional Features}
\textit{Scan resumption.} The framework supports resuming interrupted scans using the tile index written during shutdown (see section~\ref{sec:ParallelizationStrategy}). For scans resumed after sample exchange or between microscope sessions, users redefine the panorama corner points in the same order, and the framework recalculates the coordinate mapping.

\textit{Parameter interpolation.} For large panoramas, imaging parameters (working distance, brightness, contrast) can be set at the corners and linearly interpolated across the scan area to compensate for sample tilt or varying surface conditions.

\textit{Error recovery.} If any process terminates unexpectedly, the framework calls \texttt{reset\_microscope()} to return the stage to its initial position and restore all imaging parameters. Currently, the framework handles errors at the process level: if any of the two processes (SEM, or region proposal) throws an unhandled exception, the shutdown procedure is triggered.

\subsubsection{Data Output}

Each tile is saved in a separate directory within a root directory specified in the configuration file. A copy of the configuration file is stored in the root directory. Within each tile directory, images are named according to their position and extent in the field of view as \texttt{x\_y\_width\_height.tiff}. This naming convention applies to both the fast scan images and the rescanned regions, allowing each image to be located within the tile from its filename alone. The framework saves images as TIFF files with microscope metadata embedded in the file header, including acceleration voltage, probe current, and detector settings. The tile position in the microscope coordinate system is stored in a separate JSON file within the tile directory. Rescanned regions can be linked back to the full panorama by combining their position within the tile with the tile position in the panorama.

\subsubsection{Reproducibility}

The framework records all information necessary to reproduce a measurement. The configuration file, which specifies the imaging parameters, detection method and its parameters, and panorama coordinates, is saved alongside the output data. To repeat a measurement, users provide the saved configuration file to the framework without further modification. Combined with the scan resumption feature, this allows measurements to be continued across sessions or restarted from scratch with identical parameters.
\section{Case Study: Damage Characterization in Dual-Phase Steel}

\subsection{Materials and Methods}
\begin{figure}[!ht]
    \centering
    \includegraphics[width=\linewidth]{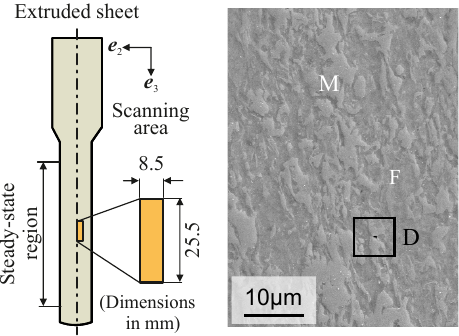}
    \caption{Schematic representation of the scanned area on the left and a representative micrograph on the right. M marks martensite, F ferrite, and D a damage site.}
    \label{fig:sheet_extrusion_specimen}
\end{figure}

Dual-phase steels are a popular subclass of advanced high strength steels often found in structural applications \cite{tasan_overview_2015}. As, such their prominence is strongly tied to their mechanical properties, especially energy absorption and component life-time, which are impacted negatively by increased damage prevalence. Dual phase steels are composed of a ferritic matrix, compounded with martensite islands, and the observed damage phenomena are usually tied to either fracture of martensite or delamination at the interface between martensite and ferrite. 
In order to validate the proposed framework for efficient damage detection, formed specimens with damage sites distributed across large flat areas are preferred. Therefore, the material is plastically formed by a so-called sheet extrusion process to generate damage sites (process details can be found in \cite{kolpak_large_2021}). The initial DP800 sheet is positioned between two billet halves made of a bulk material 16MnCrS5 with similar strength. Subsequently, the assembled workpiece is cold extruded using a triple-acting hydraulic press (SMG HZPUJ 260/160). 
This forming process offers the advantage of homogeneous deformation at large plastic strains. In addition, the extruded sheet exhibits a steady-state region due to its quasi-stationary process characteristic. Moreover, the process is symmetric with respect to the $\boldsymbol{e}_\mathrm{1}$-$\boldsymbol{e}_\mathrm{2}$ and $\boldsymbol{e}_\mathrm{2}$-$\boldsymbol{e}_\mathrm{3}$ planes. Therefore, considering half of the specimen width at any position within the steady-state region as scanning area is representative for the entire extruded part.

Samples were cut from the extruded sheets by electrical discharge machining (EDM). The cross-sections of interest were then prepared using standard metallographic procedures to obtain a flat, scratch-free surface suitable for SEM observations. For this purpose, the specimens were successively ground with SiC abrasive papers up to a grit size of 4000, followed by polishing on DAC cloths using water-based diamond suspensions with particle sizes of \SI{3}{\micro\meter} and \SI{1}{\micro\meter}. A final chemo-mechanical polishing step with an oxide polishing suspension (OPS, nominal particle size \SI{0.25}{\micro\meter}) was carried out to achieve a mirror-like finish. Subsequently, the polished surfaces were etched in \SI{2}{\percent} nital for \SI{6}{\second}, revealing the dual-phase microstructure and providing sufficient contrast to distinguish between martensitic and ferritic regions.

Once polished samples were mounted in the SEM (Tescan, Clara) and imaged with an Everhart–Thornley  secondary electron detector at a working distance of \SI{10}{\milli\meter}, an accelerating voltage of \SI{20}{kV}, and a beam current of \SI{3}{\nano\ampere}. A schematic representation of the sample and the scanned region, together with an exemplary micrograph are shown in \autoref{fig:sheet_extrusion_specimen}.

To enable systematic evaluation of detection performance, a validation dataset was acquired by recording a full panorama using the analysis imaging parameters $\theta_{\text{analysis}}$ without the automated detection and rescanning workflow. The panorama consisted of $10 \times 30$ tiles, each covering a field of view of \SI{100}{\micro\metre} $\times$ \SI{100}{\micro\metre}, with \SI{15}{\percent} overlap between adjacent tiles. Each tile was imaged twice in immediate succession with different parameter sets, minimizing drift between acquisitions. The fast scan parameters $\boldsymbol{\theta}_{\text{fast}}$ used a dwell time of \SI{32}{\micro\second} at $768 \times 768$ pixels, yielding a resolution of \SI{130}{\nano\metre} per pixel. The analysis parameters $\boldsymbol{\theta}_{\text{analysis}}$ used a dwell time of \SI{32}{\micro\second} at $3072 \times 3072$ pixels, yielding a resolution of \SI{33}{\nano\metre} per pixel. After acquiring both images at one position, the stage moved to the next tile. The total imaged area was \SI{2.1675}{\milli\metre\squared}.

\subsection{Detection Pipeline and Reference Configuration}

In dual-phase steels, damage sites manifest as localized clusters of dark pixels in scanning electron micrographs due to the reduced signal intensity at voids and cracks. To detect such features, Kusche et al.\ \cite{kusche_large-area_2019} proposed a two-stage approach: first, pixels with intensity below a threshold $\tau$ are identified as damage candidates, and then spatially coherent groups of candidates are distinguished from noise using density-based clustering (DBSCAN) \cite{ester_density-based_1996}. DBSCAN assigns each candidate pixel to a cluster if at least $n_{\text{min}}$ other candidate pixels lie within a radius of $\varepsilon$~pixels; candidates that do not meet this density criterion are discarded as noise. Each resulting cluster is considered a potential damage site.

Since the detected damage population depends on the choice of detection parameters, we evaluated framework performance across a range of configurations to ensure that the results are not specific to a single parameter set. The parameter ranges were chosen to cover a broad space of possible detection outcomes, varying both the minimum spatial extent and the required local density of damage clusters. The threshold $\tau$ defines the maximum grayscale intensity below which a pixel is classified as a damage candidate. Values of $\tau \in [45, 50, 55, 60]$ were selected based on visual inspection of the analysis-scan micrographs, spanning the range from conservative detection of only the darkest voids to more permissive inclusion of lighter damage features. For DBSCAN clustering, the clustering radius $\varepsilon$ and minimum neighbours $n_{\text{min}}$ are coupled parameters: therefore we scaled $n_{\text{min}}$ with $\varepsilon$, testing $\varepsilon = 1$ with $n_{\text{min}} \in [4, 5]$, $\varepsilon = 2$ with $n_{\text{min}} \in [10, 13]$, and $\varepsilon = 3$ with $n_{\text{min}} \in [15, 20, 25]$. 

\subsection{Matching Methodology}
The following evaluation assesses the detection performance of one specific detection pipeline used with the framework: a fast scan at reduced resolution with bicubic and Lanczos \cite{duchon_lanczos_1979} interpolation to the analysis resolution, followed by DBSCAN-based detection. The detection performance depends on the choice of $\boldsymbol{\theta}_{\text{fast}}$, upsampling method, and detection algorithm, and is independent of the framework itself.

To evaluate detection performance, we compared damage sites detected in the $\boldsymbol{\theta}_{\text{fast}}$ images against reference sites identified in the $\boldsymbol{\theta}_{\text{analysis}}$ images. To establish this correspondence, matching was performed using pixel overlap rather than centroid distance. This approach accounts for the fact that processing can alter cluster morphology: a single fragmented cluster in the high-resolution image may appear as a connected region after interpolation, or vice versa. With centroid-based matching, if two nearby clusters in the reference image are detected as a single merged cluster, only one reference cluster would be matched and the other would be counted as a false negative, even though the proposed rescan region covers both. Overlap-based matching correctly identifies both clusters as detected in this case.

The matching procedure was as follows. First, reference damage sites were identified in the $\boldsymbol{\theta}_{\text{analysis}}$ images using thresholding followed by DBSCAN clustering. For each reference cluster, we recorded the set of pixel coordinates belonging to that cluster. Second, detection candidates were identified in the $\boldsymbol{\theta}_{\text{fast}}$ images using the same algorithmic pipeline with varying threshold and clustering parameters. Third, for each reference cluster, we checked whether any detection cluster shared at least one pixel with the reference cluster. A reference site was marked as detected if such overlap existed. This permissive criterion is appropriate because the evaluation targets detection, not segmentation: any overlap means a rescan region of size $w_{\text{region}} \times h_{\text{region}}$ is placed around the detected cluster, which is sufficient to acquire an analysis quality image of the reference site. Metrics such as IoU or Dice \cite{dice_measures_1945} would additionally penalise differences in cluster shape and position between the fast and analysis images, which are not relevant for the rescanning task.

The paired images were acquired sequentially at each tile position, with the $\boldsymbol{\theta}_{\text{analysis}}$ image recorded immediately after the $\boldsymbol{\theta}_{\text{fast}}$ image. Despite this, minor stage drift between acquisitions can introduce spatial offsets between corresponding features. To account for this, we dilated each reference cluster by 10 pixels prior to overlap checking. This dilation compensates for alignment uncertainty between the two separately acquired images and does not affect the overlap criterion itself: a detection must still overlap with the dilated reference cluster to count as a match. The value of 10 pixel was chosen as a conservative upper estimate of the expected drift. Since the rescan patches are much larger than this offset, even a detection at the edge of the dilated region would still produce a rescan window that fully covers the reference site.

Detection rate was calculated as the fraction of reference sites with at least one overlapping detection:
\begin{equation}
    R_{\text{detection}} = \frac{N_{\text{overlap}}}{N_{\text{reference}}}
\end{equation}
where $N_{\text{overlap}}$ is the number of reference clusters overlapping with at least one detection cluster, and $N_{\text{reference}}$ is the total number of reference clusters.

Efficiency was defined as the fraction of the total scan area that does not require rescanning:
\begin{equation}
    E = 1 - \frac{A_{\text{rescan}}}{A_{\text{total}}}
\end{equation}
where $A_{\text{rescan}}$ is the area covered by rescan patches around detected clusters and $A_{\text{total}}$ is the total scanned area. An efficiency of \SI{100}{\percent} would indicate that no regions were rescanned, meaning no POIs were detected. In practice, efficiency is bounded from above by the theoretical maximum corresponding to perfect detection of all ground truth POIs with no false positives. This area-based metric is independent of microscope parameters such as dwell time, allowing comparison across different imaging conditions.

\subsection{Accuracy-Efficiency Trade-off}
\begin{figure*}[htbp]
    \centering
    \includegraphics[width=\linewidth]{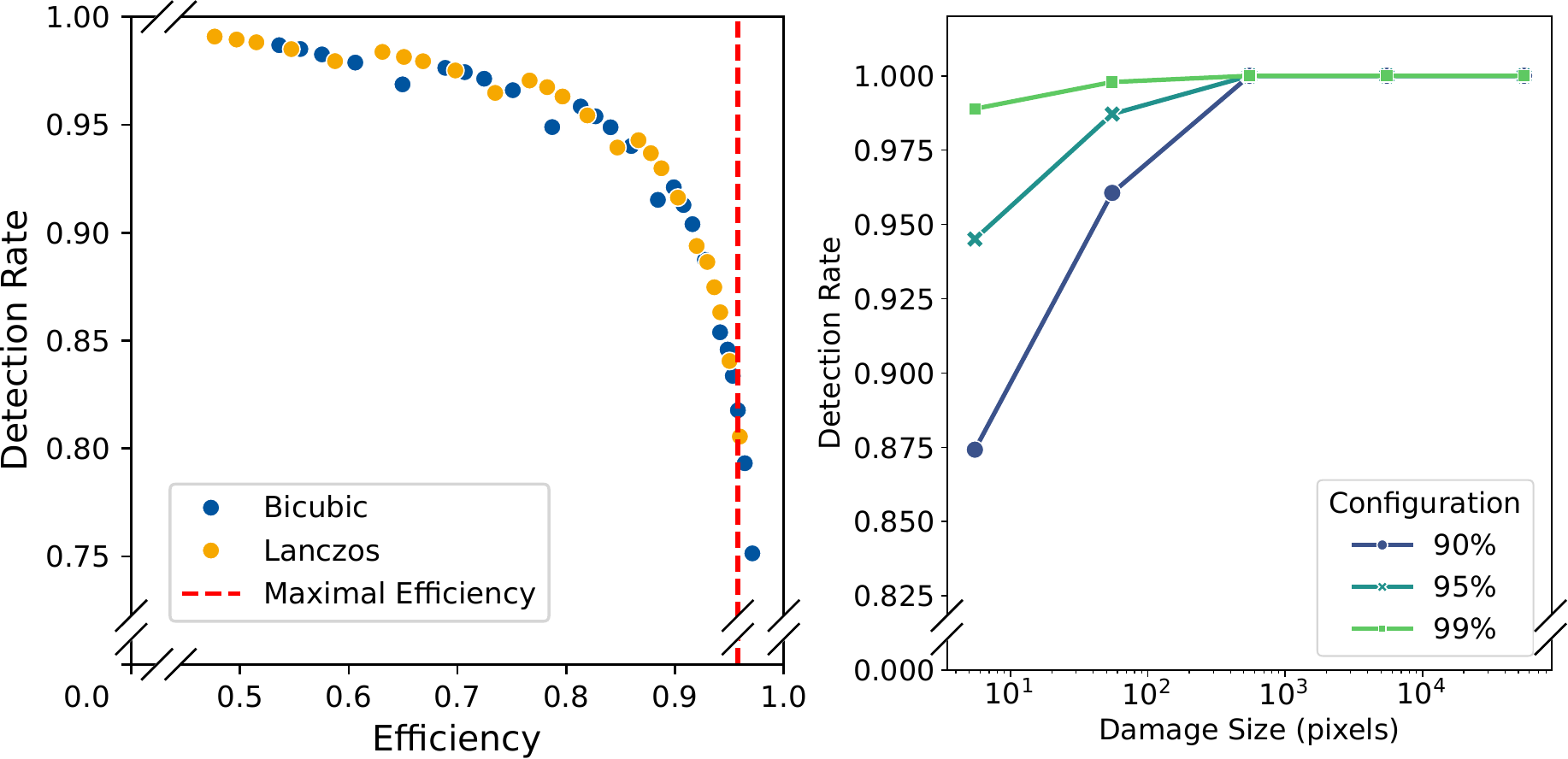}
    \caption{Detection performance for $\tau = 55$, $\varepsilon = 1$, $n_{\text{min}} = 5$. (a) Pareto front showing the trade-off between detection rate and efficiency. The dashed line indicates the theoretical maximum efficiency assuming perfect detection of all ground truth clusters. (b) Detection rate as a function of cluster size for three operating points corresponding to 99\%, 95\%, and 90\% overall detection rate. Lines connecting data points added to guide the eye.}
    \label{fig:pareto_comparison}
\end{figure*}

\autoref{fig:pareto_comparison}a shows the trade-off between detection rate and efficiency for a representative parameter configuration ($\tau = 55$, $\varepsilon = 1$, $n_{\text{min}} = 5$). 
These parameters were chosen for the detection of damage in DP800 under the specific imaging conditions used in this study. The threshold $\tau = 55$ on the 8-bit grayscale range captures the dark contrast of damage sites while excluding the ferritic matrix and depends on the imaging parameters such as brightness, contrast, and detector settings. The clustering parameters $\varepsilon = 1$ and $n_{\text{min}} = 5$ ensure sensitivity to small and elongated crack-like features. The Pareto front follows the expected trend: at the highest detection rate of \SI{99}{\percent}, efficiency is approximately \SI{48}{\percent}, meaning roughly half of the scanned area would require rescanning. Relaxing detection requirements improves efficiency substantially—at \SI{95}{\percent} detection rate, efficiency increases to \SI{87}{\percent}, and at \SI{90}{\percent} detection rate, efficiency reaches \SI{95}{\percent}. The theoretical maximum efficiency of \SI{96}{\percent}, corresponding to perfect detection of all ground truth clusters with no false positives, provides an upper bound for this configuration.

These efficiency gains translate directly into reduced acquisition time. The total acquisition time of the detection pipeline presented here, $t_{\text{SPARSE}}$ relative to that required for conventional full-area high-resolution imaging, $t_{\boldsymbol{\theta}_\text{analysis}}$ can be estimated as
\begin{equation}
    \frac{t_{\text{SPARSE}}}{t_{\boldsymbol{\theta}_\text{analysis}}} = \frac{1}{16} + \frac{A_{\text{rescan}}}{A_{\text{total}}} = \frac{17}{16} - E
\end{equation}
where the factor $1/16$ reflects the reduced pixel count of the fast scan relative to the high-resolution scan, and the second term accounts for the rescanned area. For the operating points discussed above, this corresponds to approximately \SI{58}{\percent} of the conventional acquisition time at \SI{99}{\percent} detection rate, \SI{19}{\percent} at \SI{95}{\percent}, and \SI{10}{\percent} at \SI{90}{\percent}. Even conservative operating points with near-complete detection thus yield substantial time savings, while accepting moderate reductions in detection rate can reduce acquisition time by an order of magnitude. For example, acquiring \SI{1}{mm^2} with $\boldsymbol{\theta}_{\text{analysis}}$ at a dwell time of \SI{32}{\micro \second} and a resolution of 3072 pixel per \SI{100}{\micro \metre} requires approximately \SI{12}{\hour}. At the conservative operating point of \SI{95}{\percent} detection rate, this reduces to approximately \SI{2.3}{h}, and at \SI{90}{\percent} detection rate to approximately \SI{1.2}{h}. These estimates are lower bounds based on the ratio of scanned pixels and do not account for microscope-specific overhead such as scan setup time and line flyback. Actual time savings will depend on the overhead characteristics of the specific instrument.

\autoref{fig:pareto_comparison}b shows the detection rate as a function of cluster size for three operating points along the Pareto front (\SI{99}{\percent}, \SI{95}{\percent}, and \SI{90}{\percent} overall detection rate). Detection rate generally increases with cluster size, as larger damage sites produce stronger signals in the fast scan images. Detection rate is lowest for the smallest clusters and increases with cluster size, as larger damage sites produce stronger signals in the fast scan images.

\subsection{Sensitivity to Detection Parameters}

The full evaluation across all parameter configurations is shown in \autoref{fig:full_parameter_evaluation}. The results demonstrate consistent behaviour across the tested parameter space: all configurations produce similar Pareto front shapes, with the primary difference being the range of achievable detection rates and efficiencies. Importantly, each parameter configuration defines its own reference population: the ground truth clusters are identified in the $\boldsymbol{\theta}_{\text{analysis}}$ images using the same parameters, so the detection rate reflects how well the fast scan reproduces the results of the respective full-resolution analysis. Configurations with stricter clustering requirements (larger $\varepsilon$ and $n_{\text{min}}$) target larger and denser clusters, which produce stronger signals in the fast scan images and are therefore more reliably detected.  This shifts the Pareto front toward higher detection rates and higher efficiencies, but over a smaller population of damage sites. Conversely, configurations with more permissive clustering parameters detect smaller and sparser features, resulting in lower baseline detection rates and reduced efficiency due to the larger number of rescan candidates. These results indicate that the framework performance is robust across reasonable parameter choices, with the specific configuration primarily determining which subset of the damage population is targeted.

\section{Discussion}

\subsection{Comparison to Prior Work}

Recent work by Clusiau et al.\ demonstrated workflow automation for SEM acquisitions with feature tracking for nanoparticle analysis \cite{clusiau_workflow_2025}. Our framework shares the goal of reducing acquisition time through selective high-resolution imaging but differs in two main aspects. First, our implementation parallelizes scanning, detection, and rescanning, decoupling computation time from acquisition time. In a sequential workflow, any time spent on detection directly adds to the total acquisition time, which limits users to simple and fast detection methods. With our parallel approach, more demanding algorithms can run alongside scanning without penalty, as long as processing finishes before the next partition is ready. If this constraint becomes limiting, the modular architecture allows computation to be distributed across multiple processing instances, ensuring that acquisition speed remains the bottleneck rather than detection. Second, we systematically quantify the accuracy-efficiency trade-off across detection parameters, allowing users to choose an operating point that balances detection completeness against acquisition time.

\subsection{Detection Performance in Dual-Phase Steel}

\begin{figure*}
    \centering
    \includegraphics[width=\linewidth]{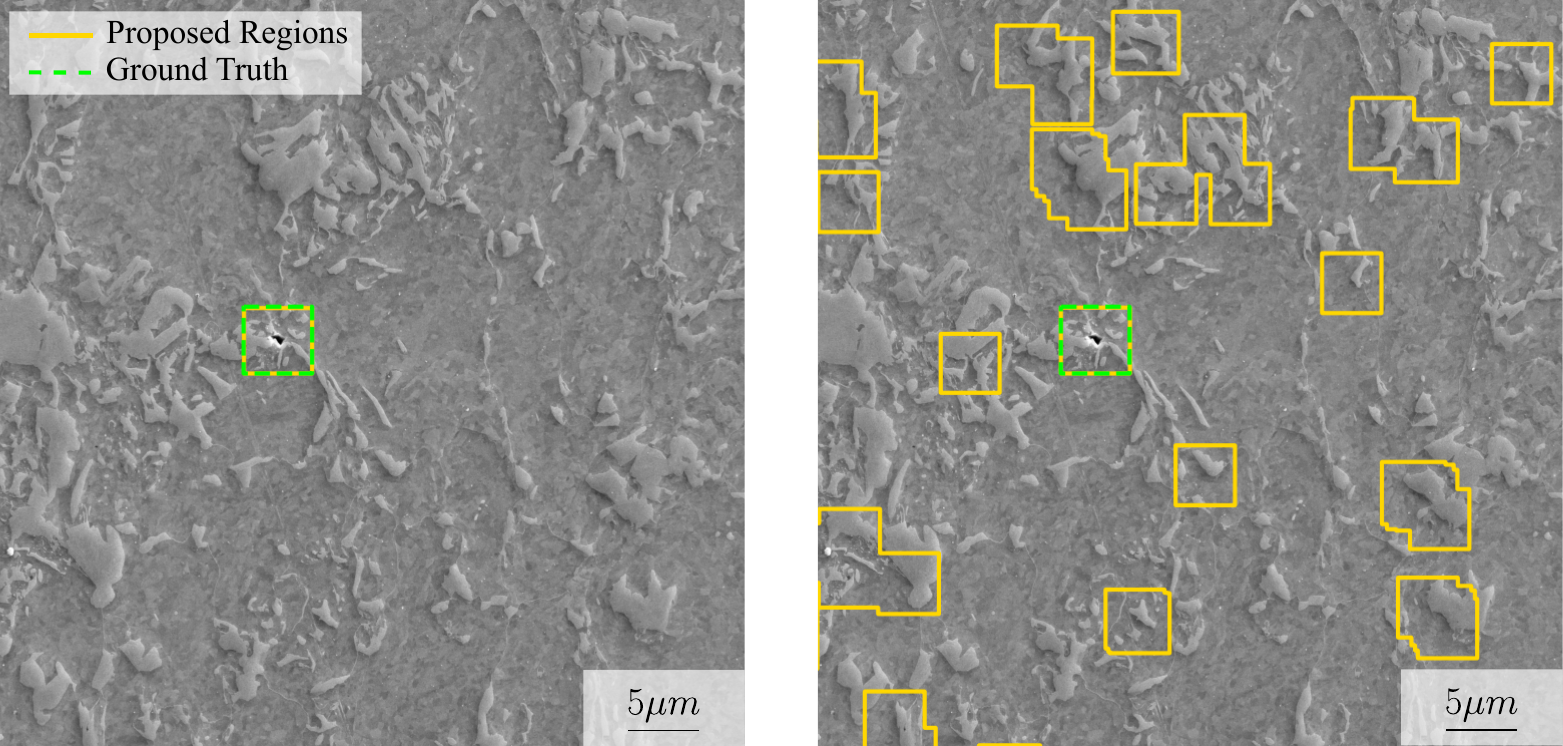}
    \caption{Comparison of rescan regions for two operating points on the Pareto front. Outlines of proposed rescan regions are shown in yellow, ground truth damage sites in green. Left: at \SI{95}{\percent} detection rate (\SI{87}{\percent} efficiency), only isolated regions require rescanning. Right: at \SI{99}{\percent} detection rate (\SI{48}{\percent} efficiency), the rescan regions cover approximately half of the shown area.}
    \label{fig:mask_comparison}
\end{figure*}

For the selected parameter configuration ($\tau = 55$, $\varepsilon = 1$, $n_{\text{min}} = 5$), nearly all damage site candidates could be identified, achieving a detection rate of \SI{99}{\percent}. However, this high detection rate comes at a significant cost: the rescan mask covers approximately half of the total area. For microscopes without the capability to scan arbitrary masked regions, this would require rescanning large bounding boxes, potentially negating the time savings. \autoref{fig:mask_comparison} illustrates this trade-off: at \SI{99}{\percent} detection rate, the surface is densely covered with rescan patches, while at \SI{95}{\percent} detection rate, only isolated regions require rescanning.

Furthermore, detected POIs are not necessarily damage sites, as some correspond to imaging artifacts such as shadows or contrast variations. The true number of damage sites is smaller, as some detections correspond to imaging artifacts such as shadows or contrast variations. This is a known challenge in damage characterisation of dual-phase steels: Medghalchi et al. \cite{medghalchi_damage_2020} addressed this by including a shadow class alongside the damage mechanism classes in the classification task. In principle, the framework supports a workflow where detected candidates are first rescanned at analysis quality and then classified using a trained model that distinguishes damage mechanisms from artifacts. However, implementing and evaluating such a classification pipeline is beyond the scope of this work.

The chosen DBSCAN configuration ($\varepsilon = 1$, $n_{\text{min}} = 5$) targets small, dense clusters and represents a challenging case for interpolation-based detection. We deliberately selected this configuration for the main analysis because it defines the a demanding yet realistic operating regime: if the framework performs well for these small, difficult-to-detect features, performance for larger damage sites is expected to be at least as good. As shown in Figure S1, this expectation is confirmed: other physically reasonable configurations achieve substantially better performance. For example, with $\tau = 55$, $\varepsilon = 2$, and $n_{\text{min}} = 13$, the detection rate reaches \SI{100}{\percent} with efficiency above \SI{90}{\percent}, corresponding to an estimated acquisition time of approximately \SI{16}{\percent} of the conventional full-area high-resolution approach.

The practical value of simple interpolation-based detection depends on the microscope's scanning capabilities. When mask-based scanning is available, the framework achieves substantial time savings even with moderate false positive rates, as only the masked pixels are rescanned. In this case, simple interpolation methods suffice for identifying most ground truth damage sites while still accelerating acquisition. However, when only bounding box or greedy merging strategies are available, false positives have a larger impact on efficiency: each false positive enlarges the rescan region, and at high detection rates the bounding boxes may cover most of the tile area (\autoref{fig:mask_comparison}). In such cases, detection methods that reduce false positives while maintaining high detection rates become essential. Learned approaches may provide the necessary improvement in precision to make the framework practical on hardware with more limited scanning flexibility. For example, in previous work \cite{reclik_resolution_2025}, we demonstrated that a reference-based super-resolution network trained on dual-phase steel micrographs outperforms standard interpolation in preserving microstructural features relevant to damage detection, such as phase boundaries and voids. Due to the modular detection interface, integrating such a method requires only replacing the detection function, without modifications to the scanning workflow or parallelization strategy. This could improve damage identification in the enhanced fast-scan images, reducing false positives while maintaining high detection rates.

\subsection{Implications for Statistical Analysis of Rare Events}

Statistically meaningful analysis of rare microstructural features requires scanning areas far larger than typically feasible with conventional high-resolution imaging. In dual-phase steels, previous work has established methods for large-area damage characterisation and classification \cite{kusche_large-area_2019, medghalchi_damage_2020}, three-dimensional damage analysis \cite{medghalchi_three-dimensional_2023}, and resolution enhancement \cite{reclik_resolution_2025}. However, the acquisition time required for high-resolution panoramic imaging remains a key bottleneck that limits the areas that can be practically examined.

The framework presented here directly addresses this bottleneck by substantially reducing the acquisition time per unit area. Crucially, these time savings can be reinvested: at \SI{10}{\percent} of the conventional acquisition time: approximately ten times the area can be scanned in the same time budget or an in-situ straining experiment enabled that scans the same area at several strain increments rather than relying on a single post-mortem analysis. The importance of large-area analysis has been demonstrated by Medghalchi et al.\ \cite{medghalchi_automated_2024}, who showed that martensite phase fraction measurements in the same DP800 steel exhibit large scatter even for areas exceeding \SI{1}{\milli\metre\squared}, due to long-range microstructural heterogeneity such as martensite banding. Their analysis revealed that window sizes commonly used in the literature are insufficient for representative measurements, and that the associated uncertainty often exceeds the precision at which phase fractions are reported. While their work focused on phase fraction, the same argument applies to damage statistics: reliable characterisation of spatially heterogeneous features requires surveying areas substantially larger than current practice, which is only feasible if the acquisition time per unit area can be significantly reduced.

For the dual-phase DP800 steel studied here, the choice of operating point can be guided by the downstream analysis requirements. At \SI{99}{\percent} detection rate, nearly all damage candidates are captured, but acquisition still requires approximately \SI{58}{\percent} of the conventional imaging time. This operating point is appropriate when completeness is critical, for example when characterising the full damage size distribution including the smallest feature. At \SI{95}{\percent} detection rate, acquisition time drops to approximately \SI{19}{\percent} of the conventional approach. The missed detections are predominantly the smallest clusters (\autoref{fig:pareto_comparison}b), which contribute least to the overall damage area and are most likely to include false positives. When applying a minimum cluster size of \SI{20}{} pixel for damage classification in DP800, only three candidates are missed at this operating point, none of which correspond to actual damage sites upon manual inspection. For analyses focused on damage morphology, spatial distribution, or correlation with microstructural features, this represents a practical sweet spot, the statistical power gained from scanning larger areas more than compensates for the loss of a small number of marginal detections. At \SI{90}{\percent} detection rate, acquisition time is reduced to approximately \SI{10}{\percent}, enabling rapid surveys of very large areas at the cost of missing some smaller damage sites.

The acceptability of missing small damage candidates depends on the analysis objective. The absolute lower bound on detectable damage size is set by the analysis resolution, which is chosen to resolve the mechanisms underlying damage nucleation. Beyond this limit, the physically relevant size range is constrained by the application: unless an investigation of nucleation mechanisms is the specific objective, those damage sites may considered most important that are closer to becoming critical by growth and coalescence and therefore affect the macroscopic failure behaviour of the material, while very small sites must first reach a certain size to become mechanically relevant. As long as the chosen detection threshold is applied consistently across experiments, quantitative comparisons remain valid regardless of where exactly this threshold is set.

\subsection{Practical Recommendations}

Based on the evaluation presented here, several observations may inform users applying the framework to new materials and feature types:

\begin{enumerate}
    \item \textit{Simple methods as a baseline.} Interpolation-based detection requires no training data and provided a strong baseline in this study. The achievable efficiency depends on the material, microstructure, and feature size, and should be evaluated before investing in more complex methods.
    \item \textit{Paired validation data.} Acquiring a small paired dataset with both $\boldsymbol{\theta}_{\text{fast}}$ and $\boldsymbol{\theta}_{\text{analysis}}$ images enables quantification of the detection performance for a given combination of material, imaging parameters, and detection method. This allows informed selection of operating points on the Pareto front.
    \item \textit{Feature size relative to resolution.} In this study, features near the resolution limit of the fast scan showed reduced detection rates. Users working with similarly small features may need to adjust the resolution ratio or detection method accordingly.
    \item \textit{Detection rate over efficiency.} In the context of this framework, detecting regions that do not contain features of interest only cost additional acquisition time, while missing actual features leads to incomplete data. Optimising for high detection rate at the cost of reduced efficiency is therefore generally preferable.
\end{enumerate}

\section{Conclusion}
We presented an open-source Python framework for efficient large-area SEM characterisation of rare microstructural features. This SPARSE framework decouples detection from acquisition through a generic microscope interface that separates microscope-specific control from the scanning logic. This design is intended to allow deployment across different microscope platforms without modifications to the core framework. Scanning, detection, and rescanning are parallelized using separate processes, ensuring that computation time is hidden behind acquisition time. Users can select their operating point on the accuracy-efficiency trade-off based on application requirements.

Validation on damage detection in dual-phase DP800 steel demonstrated that simple interpolation-based detection achieves substantial time savings without requiring training data. Detection rates above \SI{95}{\percent} were maintained while reducing acquisition time to approximately \SI{19}{\percent} of the conventional full-area high-resolution approach based on the ratio of
scanned pixels. For the specific damage characteristics and density observed in this DP800 steel at the given degree of deformation, and the minimum cluster size established by \cite{kusche_large-area_2019} for damage classification, no actual damage sites are missed at this operating point. For stricter clustering parameters, which target larger and more prominent damage sites, all ground truth candidates were detected while reducing acquisition time to approximately \SI{16}{\percent}. The trade-off behaviour was robust across a wide range of detection parameters, with stricter configurations identifying smaller features at the cost of reduced efficiency. These time savings can be reinvested into scanning larger areas, directly improving the statistical power of rare-event analyses.

The detection method presented here serves as a baseline to validate the framework. The modular detection interface allows it to be substituted with different approaches, such as resolution enhancement \cite{reclik_resolution_2025}, localisation directly in the initial images, or denoising, without modifications to the scanning workflow. The framework also provides practical features for extended imaging campaigns, such as scan resumption across sessions, parameter interpolation, and failure recovery.

The complete implementation will be made available upon publication and upon request during peer-review.

\section{Acknowledgements} 
The authors express their gratitude to the Deutsche Forschungsgemeinschaft (DFG) for financial support in context of the Collaborative Research Centre CRC/Transregio 188/2 “Damage Controlled Forming Processes”, project no. 278868966, and the Collaborative Research Centre CRC 1394 "Structural and chemical atomic complexity– from defect phase diagrams to material properties" , project no. 409476157. This project has also received funding from the European Research Council (ERC) under the European Union’s Horizon 2020 Research and Innovation Programme (Grant Agreement No. 101168203 "TailorPlast").

The authors thank TESCAN GROUP a.s. for their support and collaboration.

\section*{Author Contribution Statement}

\textbf{Tom Reclik:} Software, Methodology, Formal analysis, Investigation, Writing - Original Draft
\textbf{Jan Gerlach:} Methodology, Investigation,  Writing - Review \& Editing
\textbf{Maximilian A. Wollenweber:} Methodology, Investigation,  Writing - Review \& Editing
\textbf{Yannis P. Korkolis:} Conceptualisation, Writing - Review \& Editing, Supervision, Funding Acquisition
\textbf{Sandra Korte-Kerzel:} Conceptualisation, Writing - Review \& Editing, Supervision, Funding Acquisition
\textbf{Ulrich Kerzel:} Conceptualisation, Writing - Review \& Editing, Supervision, Funding Acquisition

\section*{Declaration of generative AI and AI-assisted technologies in the manuscript preparation process}

During the preparation of this work the authors used the LLMs RWTHGPT and Claude (claude.ai) in order to improve language and readability. After using this tool/service, the authors reviewed and edited the content as needed and take full responsibility for the content of the published article.

\bibliographystyle{elsarticle-num}
\bibliography{
    references.bib
}

@article{kusche_large-area_2019,
	title = {Large-area, high-resolution characterisation and classification of damage mechanisms in dual-phase steel using deep learning},
	volume = {14},
	issn = {1932-6203},
	doi = {10.1371/journal.pone.0216493},
	language = {en},
	number = {5},
	urldate = {2023-09-06},
	journal = {PLOS ONE},
	publisher = {Public Library of Science},
	author = {Kusche, Carl and Reclik, Tom and Freund, Martina and Al-Samman, Talal and Kerzel, Ulrich and Korte-Kerzel, Sandra},
	month = may,
	year = {2019},
	pages = {e0216493},
}

@article{medghalchi_three-dimensional_2023,
	title = {Three-dimensional characterisation of deformation-induced damage in dual phase steel using deep learning},
	volume = {232},
	issn = {0264-1275},
	doi = {10.1016/j.matdes.2023.112108},
	urldate = {2023-09-21},
	journal = {Materials \& Design},
	author = {Medghalchi, Setareh and Karimi, Ehsan and Lee, Sang-Hyeok and Berkels, Benjamin and Kerzel, Ulrich and Korte-Kerzel, Sandra},
	month = aug,
	year = {2023},
	pages = {112108},
}

@article{duchon_lanczos_1979,
	chapter = {Journal of Applied Meteorology and Climatology},
	title = {Lanczos {Filtering} in {One} and {Two} {Dimensions}},
	volume = {18},
	issn = {1520-0450},
	doi = {10.1175/1520-0450(1979)018<1016:LFIOAT>2.0.CO;2},
	language = {EN},
	number = {8},
	urldate = {2024-05-02},
	journal = {Journal of Applied Meteorology and Climatology},
	publisher = {American Meteorological Society},
	author = {Duchon, Claude E.},
	month = aug,
	year = {1979},
	pages = {1016--1022},
}

@article{wollenweber_automated_2023,
	title = {On the automated characterisation of inclusion-induced damage in {16MnCrS5} case-hardening steel},
	volume = {7},
	issn = {2666-9129},
	doi = {10.1016/j.aime.2023.100123},
	urldate = {2024-07-08},
	journal = {Advances in Industrial and Manufacturing Engineering},
	author = {Wollenweber, Maximilian A. and Kusche, Carl F. and Al-Samman, Talal and Korte-Kerzel, Sandra},
	month = nov,
	year = {2023},
	pages = {100123},
}

@article{medghalchi_automated_2024,
	title = {Automated segmentation of large image datasets using artificial intelligence for microstructure characterisation and damage analysis},
	volume = {243},
	issn = {0264-1275},
	doi = {10.1016/j.matdes.2024.113031},
	urldate = {2024-10-08},
	journal = {Materials \& Design},
	author = {Medghalchi, Setareh and Kortmann, Joscha and Lee, Sang-Hyeok and Karimi, Ehsan and Kerzel, Ulrich and Korte-Kerzel, Sandra},
	month = jul,
	year = {2024},
	pages = {113031},
}

@article{medghalchi_damage_2020,
	title = {Damage {Analysis} in {Dual}-{Phase} {Steel} {Using} {Deep} {Learning}: {Transfer} from {Uniaxial} to {Biaxial} {Straining} {Conditions} by {Image} {Data} {Augmentation}},
	volume = {72},
	issn = {1543-1851},
	shorttitle = {Damage {Analysis} in {Dual}-{Phase} {Steel} {Using} {Deep} {Learning}},
	doi = {10.1007/s11837-020-04404-0},
	language = {en},
	number = {12},
	urldate = {2024-11-05},
	journal = {JOM},
	author = {Medghalchi, Setareh and Kusche, Carl F. and Karimi, Ehsan and Kerzel, Ulrich and Korte-Kerzel, Sandra},
	month = dec,
	year = {2020},
	note = {TLDR: Overall, the network performance could be greatly improved and an analysis of changes in damage behavior, here the martensite crack angle distribution, with stress state can now be performed.},
	pages = {4420--4430},
}

@article{tekkaya_forming-induced_2017,
	title = {Forming-induced damage and its effects on product properties},
	volume = {66},
	issn = {0007-8506},
	doi = {10.1016/j.cirp.2017.04.113},
	number = {1},
	urldate = {2024-12-02},
	journal = {CIRP Annals},
	author = {Tekkaya, A. Erman and Ben Khalifa, Nooman and Hering, Oliver and Meya, Rickmer and Myslicki, Sebastian and Walther, Frank},
	month = jan,
	year = {2017},
	pages = {281--284},
}

@article{tasan_overview_2015,
	title = {An {Overview} of {Dual}-{Phase} {Steels}: {Advances} in {Microstructure}-{Oriented} {Processing} and {Micromechanically} {Guided} {Design}},
	volume = {45},
	issn = {1531-7331, 1545-4118},
	shorttitle = {An {Overview} of {Dual}-{Phase} {Steels}},
	doi = {10.1146/annurev-matsci-070214-021103},
	language = {en},
	number = {Volume 45, 2015},
	urldate = {2024-12-02},
	journal = {Annual Review of Materials Research},
	publisher = {Annual Reviews},
	author = {Tasan, C. C. and Diehl, M. and Yan, D. and Bechtold, M. and Roters, F. and Schemmann, L. and Zheng, C. and Peranio, N. and Ponge, D. and Koyama, M. and Tsuzaki, K. and Raabe, D.},
	month = jul,
	year = {2015},
	pages = {391--431},
}

@article{dahmen_feature_2016,
	title = {Feature {Adaptive} {Sampling} for {Scanning} {Electron} {Microscopy}},
	volume = {6},
	copyright = {2016 The Author(s)},
	issn = {2045-2322},
	doi = {10.1038/srep25350},
	language = {en},
	number = {1},
	urldate = {2025-06-29},
	journal = {Scientific Reports},
	publisher = {Nature Publishing Group},
	author = {Dahmen, Tim and Engstler, Michael and Pauly, Christoph and Trampert, Patrick and de Jonge, Niels and Mücklich, Frank and Slusallek, Philipp},
	month = may,
	year = {2016},
	pages = {25350},
}

@article{clusiau_workflow_2025,
	title = {Workflow automation of {SEM} acquisitions and feature tracking},
	volume = {269},
	issn = {0304-3991},
	doi = {10.1016/j.ultramic.2024.114093},
	urldate = {2026-01-14},
	journal = {Ultramicroscopy},
	author = {Clusiau, Sabrina and Piché, Nicolas and Brodusch, Nicolas and Strauss, Mike and Gauvin, Raynald},
	month = mar,
	year = {2025},
	pages = {114093},
}

@article{atkinson_characterization_2003,
	title = {Characterization of inclusions in clean steels: a review including the statistics of extremes methods},
	volume = {48},
	issn = {0079-6425},
	shorttitle = {Characterization of inclusions in clean steels},
	doi = {10.1016/S0079-6425(02)00014-2},
	number = {5},
	urldate = {2026-02-23},
	journal = {Progress in Materials Science},
	author = {Atkinson, H. V. and Shi, G.},
	month = jan,
	year = {2003},
	pages = {457--520},
}

@inproceedings{ester_density-based_1996,
	address = {Portland, Oregon},
	series = {{KDD}'96},
	title = {A density-based algorithm for discovering clusters in large spatial databases with noise},
	urldate = {2026-02-23},
	booktitle = {Proceedings of the {Second} {International} {Conference} on {Knowledge} {Discovery} and {Data} {Mining}},
	publisher = {AAAI Press},
	author = {Ester, Martin and Kriegel, Hans-Peter and Sander, Jörg and Xu, Xiaowei},
	month = aug,
	year = {1996},
	pages = {226--231},
}

@article{dice_measures_1945,
	title = {Measures of the {Amount} of {Ecologic} {Association} {Between} {Species}},
	volume = {26},
	copyright = {© 1945 by the Ecological Society of America},
	issn = {1939-9170},
	doi = {10.2307/1932409},
	language = {en},
	number = {3},
	urldate = {2026-03-07},
	journal = {Ecology},
	author = {Dice, Lee R.},
	year = {1945},
	note = {\_eprint: https://esajournals.onlinelibrary.wiley.com/doi/pdf/10.2307/1932409},
	pages = {297--302},
}

@article{reclik_resolution_2025,
	title = {Resolution enhancement of scanning electron micrographs using artificial intelligence},
	volume = {253},
	issn = {0264-1275},
	doi = {10.1016/j.matdes.2025.113955},
	urldate = {2026-03-05},
	journal = {Materials \& Design},
	author = {Reclik, T. and Medghalchi, S. and Schumacher, P. and Wollenweber, M. A. and Al-Samman, T. and Korte-Kerzel, S. and Kerzel, U.},
	month = may,
	year = {2025},
	pages = {113955},
}

@article{shah_automated_2023,
	title = {Automated image segmentation of scanning electron microscopy images of graphene using {U}-{Net} {Neural} {Network}},
	volume = {35},
	issn = {2352-4928},
	doi = {10.1016/j.mtcomm.2023.106127},
	urldate = {2026-03-16},
	journal = {Materials Today Communications},
	author = {Shah, Aagam and Schiller, Joshua A. and Ramos, Isiah and Serrano, James and Adams, Darren K. and Tawfick, Sameh and Ertekin, Elif},
	month = jun,
	year = {2023},
	pages = {106127},
}

@article{zhou_efficientand_2024,
	title = {Efficientand {Robust} {Automated} {Segmentation} of {Nanoparticles} and {Aggregates} from {Transmission} {Electron} {Microscopy} {Images} with {Highly} {Complex} {Backgrounds}},
	volume = {14},
	issn = {2079-4991},
	doi = {10.3390/nano14141169},
	number = {14},
	urldate = {2026-03-16},
	journal = {Nanomaterials},
	author = {Zhou, Lishi and Wen, Haotian and Kuschnerus, Inga C. and Chang, Shery L. Y.},
	month = jul,
	year = {2024},
	pages = {1169},
}

@article{hering_damage-induced_2020,
	title = {Damage-induced performance variations of cold forged parts},
	volume = {279},
	issn = {0924-0136},
	doi = {10.1016/j.jmatprotec.2019.116556},
	urldate = {2026-04-02},
	journal = {Journal of Materials Processing Technology},
	author = {Hering, Oliver and Tekkaya, A. Erman},
	month = may,
	year = {2020},
	pages = {116556},
}

@article{hannard_characterization_2016,
	title = {Characterization and micromechanical modelling of microstructural heterogeneity effects on ductile fracture of 6xxx aluminium alloys},
	volume = {103},
	issn = {1359-6454},
	doi = {10.1016/j.actamat.2015.10.008},
	urldate = {2026-04-02},
	journal = {Acta Materialia},
	author = {Hannard, F. and Pardoen, T. and Maire, E. and Le Bourlot, C. and Mokso, R. and Simar, A.},
	month = jan,
	year = {2016},
	pages = {558--572},
}

@article{avramovic-cingara_effect_2009,
	title = {Effect of martensite distribution on damage behaviour in {DP600} dual phase steels},
	volume = {516},
	issn = {0921-5093},
	doi = {10.1016/j.msea.2009.03.055},
	number = {1},
	urldate = {2026-04-02},
	journal = {Materials Science and Engineering: A},
	author = {Avramovic-Cingara, G. and Ososkov, Y. and Jain, M. K. and Wilkinson, D. S.},
	month = aug,
	year = {2009},
	pages = {7--16},
}

@article{asik_microscopic_2019,
	title = {Microscopic investigation of damage mechanisms and anisotropic evolution of damage in {DP600}},
	volume = {739},
	issn = {0921-5093},
	doi = {10.1016/j.msea.2018.10.018},
	urldate = {2026-04-02},
	journal = {Materials Science and Engineering: A},
	author = {Aşık, E. E. and Perdahcıoğlu, E. S. and van den Boogaard, A. H.},
	month = jan,
	year = {2019},
	pages = {348--356},
}

@article{rocklin_dask_2015,
	title = {Dask: {Parallel} {Computation} with {Blocked} algorithms and {Task} {Scheduling}},
	shorttitle = {Dask},
	doi = {10.25080/Majora-7b98e3ed-013},
	language = {en},
	urldate = {2026-04-02},
	journal = {SciPy 2015},
	author = {Rocklin, Matthew},
	month = jun,
	year = {2015},
}

@article{kolpak_large_2021,
	title = {Large strain flow curves of sheet metals by sheet extrusion},
	volume = {70},
	issn = {0007-8506},
	doi = {10.1016/j.cirp.2021.03.023},
	number = {1},
	urldate = {2026-04-02},
	journal = {CIRP Annals},
	author = {Kolpak, Felix and Traphöner, Heinrich and Hering, Oliver and Tekkaya, A. Erman},
	month = jan,
	year = {2021},
	pages = {247--250},
}

\clearpage
\appendix
\onecolumn
\setcounter{figure}{0}

\section{}

\begin{figure*}[htbp!]
    \centering
    \includegraphics[trim={0 3.8cm 0 4.5cm}, clip,width=0.9\linewidth]{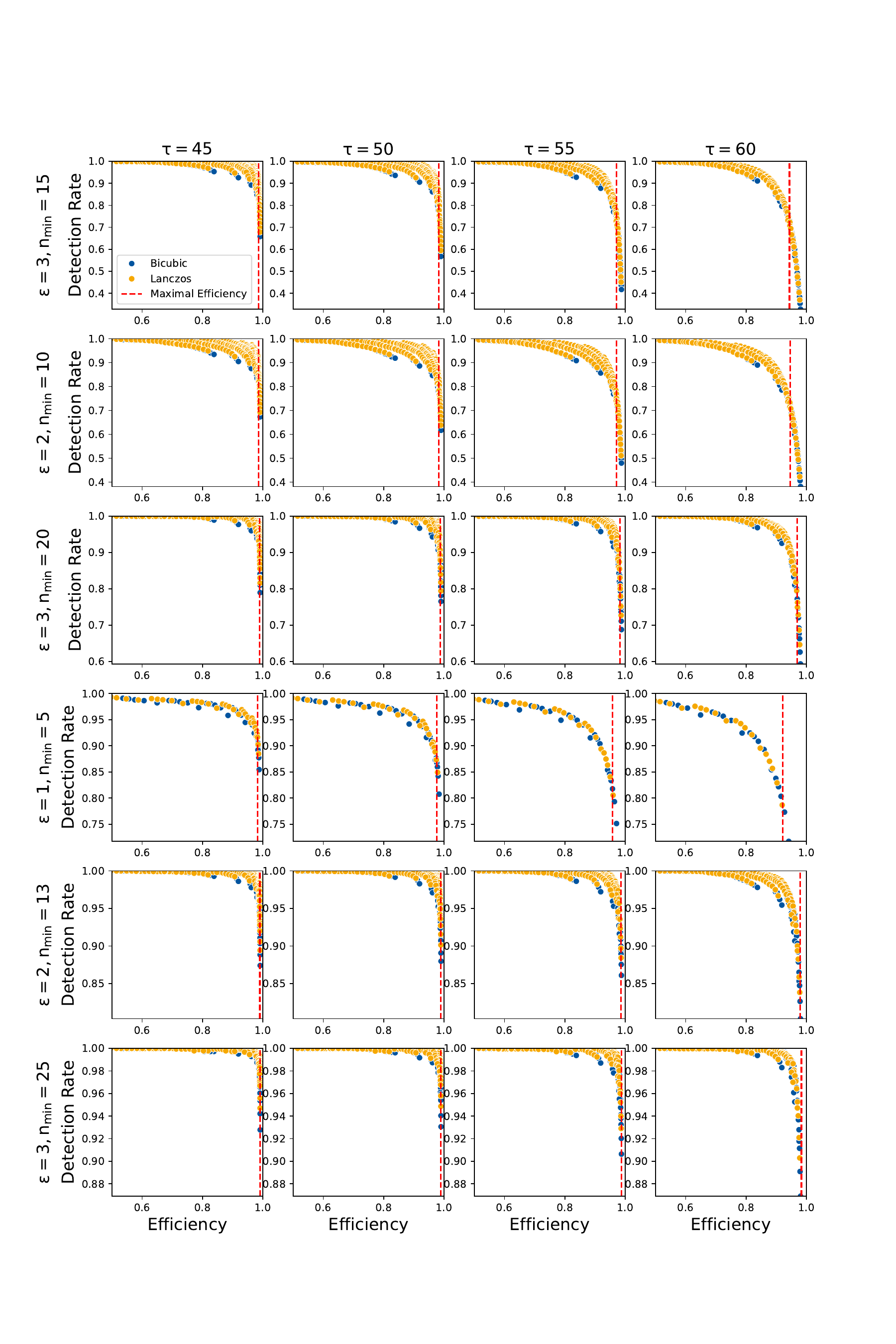}
    \caption{Detection rate versus efficiency for all tested parameter configurations. Columns correspond to intensity thresholds $\tau \in [45, 50, 55, 60]$, rows to DBSCAN configurations ordered by increasing cluster density requirements (top: $\varepsilon=3$, $n_{\text{min}}=15$; bottom: $\varepsilon=3$, $n_{\text{min}}=25$). Bicubic and Lanczos interpolation methods yield nearly identical results. The dashed red line indicates the theoretical maximum efficiency, corresponding to perfect detection of all ground truth clusters with no false positives. Stricter clustering parameters (bottom rows) yield higher detection rates but identify fewer total clusters, reducing potential time savings.}
    \label{fig:full_parameter_evaluation}
\end{figure*}

\end{document}